\documentclass{emulateapj}
\usepackage{epsfig}
\usepackage{paralist}

\begin{document}

\shortauthors{Moffat \& Toth} \shorttitle{SDSS satellite galaxies and modified gravity}

\title{Satellite galaxy velocity dispersions in the SDSS and modified gravity models}

\author{J. W. Moffat$^{\dag *}$ and V. T. Toth$^{\dag}$\\~\\}

\affil{$^\dag$Perimeter Institute, 31 Caroline St North, Waterloo, Ontario N2L 2Y5, Canada}
\affil{$^*$Department of Physics, University of Waterloo, Waterloo, Ontario N2L 3G1, Canada}

\begin{abstract}
The Sloan Digital Sky Survey (SDSS) provides data on several hundred thousand galaxies. Precise location of these galaxies in the sky, along with information about their luminosities and line-of-sight (Doppler) velocities allows one to construct a three-dimensional map of their location and estimate their line-of-sight velocity dispersion. This information, in principle, allows one to test dynamical gravity models, specifically models of satellite galaxy velocity dispersions near massive hosts. A key difficulty is the separation of true satellites from interlopers. We sidestep this problem by not attempting to derive satellite galaxy velocity dispersions from the data, but instead incorporate an interloper background into the mathematical models and compare the result to the actual data. We find that due to the presence of interlopers, it is not possible to exclude several gravitational theories on the basis of the SDSS data.
\end{abstract}


\maketitle

\section{Introduction}
\label{sec:intro}

Recently, \cite{Prada2007} presented an analysis of galaxy observations of the Sloan Digital Sky Survey (SDSS, {\tt http://www.sdss.org/}) to test gravity and dark matter in the peripheral parts of galaxies at distances 50--400~kpc from the centers of galaxies. This field of extragalactic astronomy provides one of the main arguments for the presence of dark matter \citep{ZW1994,Prada2003}.

The analysis of \cite{Prada2007} begins with identifying candidate host galaxies and candidate satellite galaxies based on their relative positions in the sky, relative velocities, and relative luminosities. After a candidate population of hosts and satellites has been identified, an {\it ad-hoc} mathematical model is used to separate the (presumed constant) background of interlopers from actual satellites. This mathematical model yields a velocity dispersion profile for the presumed satellites that is then checked against theory.

In the present work we propose an alternative approach that altogether avoids the difficult issue of identifying satellites vs. interlopers. Rather than attempting to subtract the interloper population from the data in order to construct a data set that is then hoped to represent the satellite population correctly, we endeavor to model the actual data instead, by adding an interloper population to the velocity dispersions predicted by various gravity theories. Crudely put, we extend the theory to model the data correctly, rather than massaging the data to fit within the constraints of a limited model.

In the first section of our paper, we offer a detailed description of our data analysis. In the second part, we model the data using three gravity theories. In addition to Newtonian gravity without exotic dark matter and Modified Newtonian Dynamics (MOND, \cite{Milgrom1984}), of particular interest to us is our Modified Gravity Theory (MOG, \cite{Moffat2006a,Moffat2007e}), which has been used successfully in the past to explain galaxy rotation curves \citep{Brownstein2006a}, galaxy cluster mass profiles \citep{Brownstein2006b}, cosmological observations \citep{Moffat2007c}, and gravitational lensing in the Bullet Cluster \citep{Brownstein2007} without assuming the presence of nonbaryonic dark matter. In the third section, we combine our satellite velocity dispersion predictions with the observed interloper background, and contrast the resulting predictions, as well as the cold dark matter (CDM) prediction of \cite{Prada2007} with the SDSS data. Our conclusion is that the SDSS galaxy data cannot be used to exclude any of these gravitational theories, not unless an independent, nonstatistical method is found that can be used reliably to identify individual interlopers.

\section{Data Analysis}

\begin{figure*}
 \includegraphics[width=0.48\linewidth]{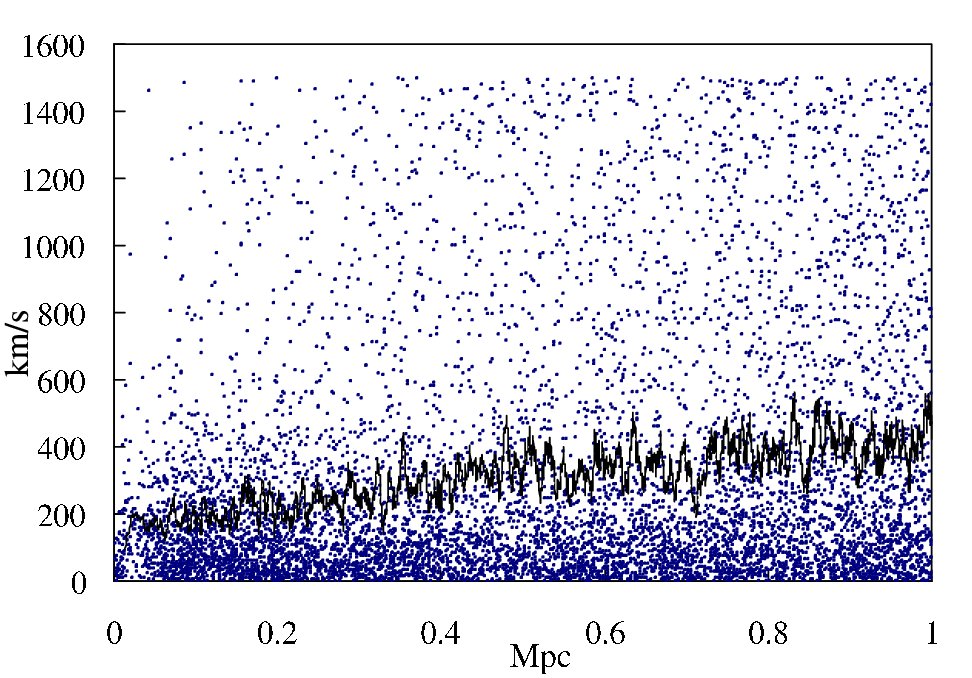}
\hskip 0.01\linewidth
\includegraphics[width=0.48\linewidth]{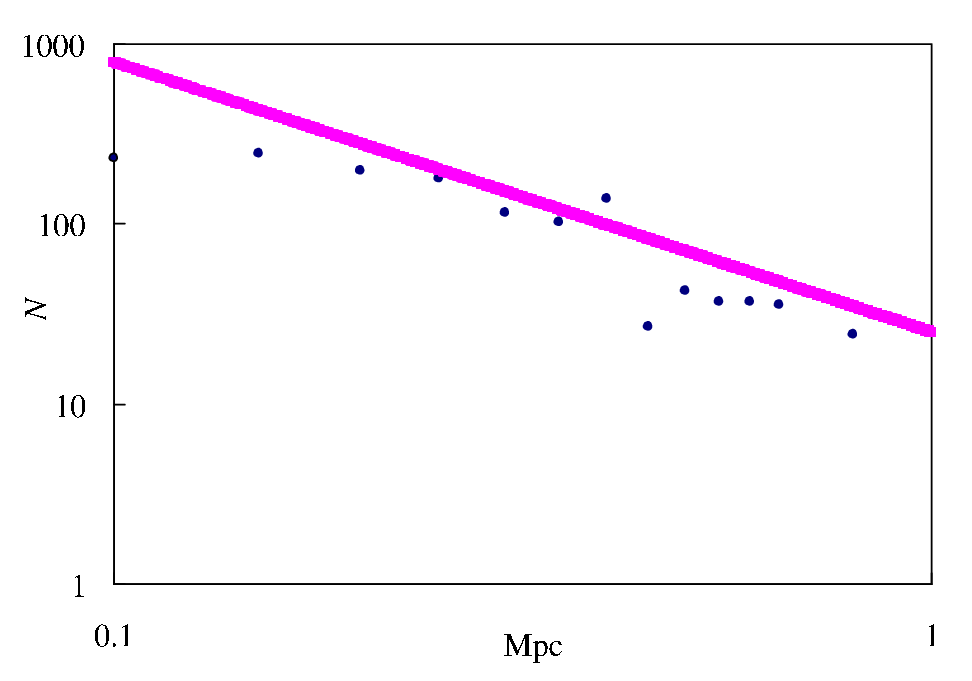}
\\
 \includegraphics[width=0.48\linewidth]{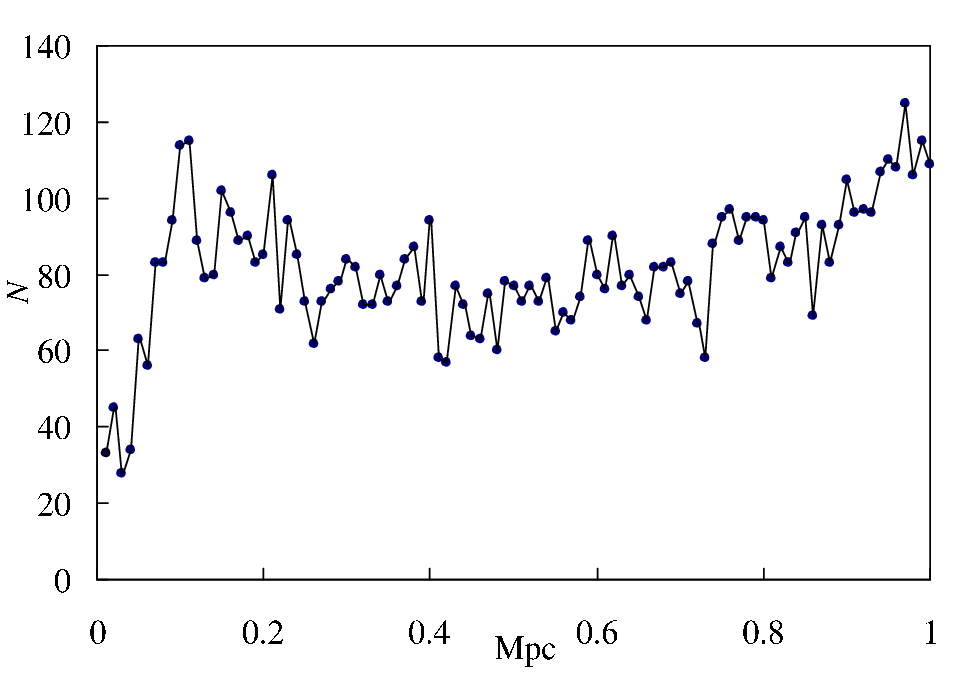}
\hskip 0.01\linewidth
\includegraphics[width=0.48\linewidth]{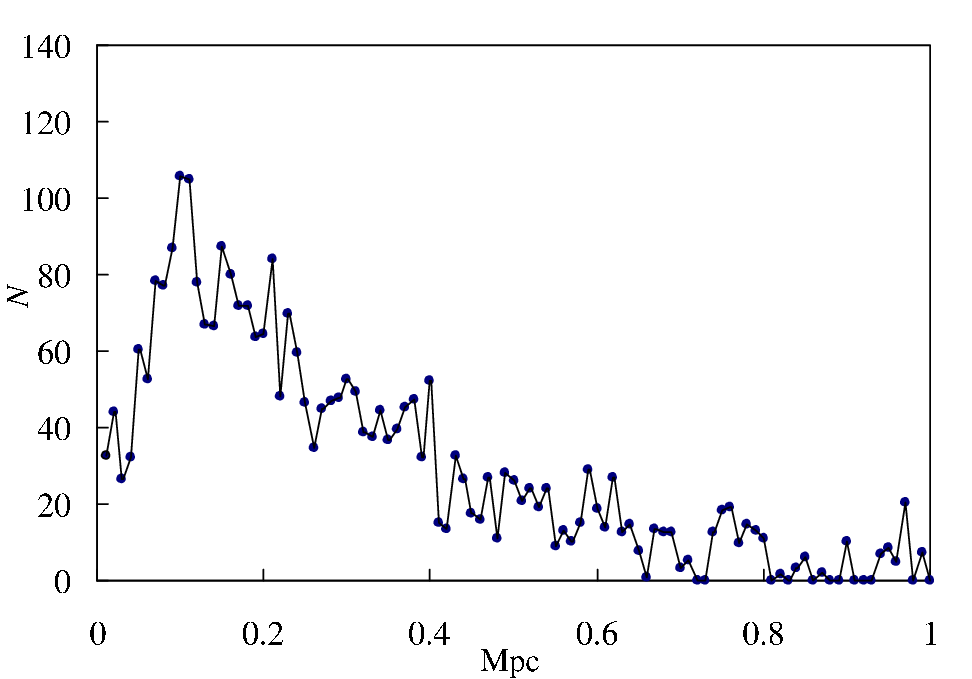}
\caption{Line-of-sight velocities (top left) and number densities (bottom left) of candidate satellite galaxies (full sample) as a function of projected distance from the candidate host. After removal of candidate interlopers, the radial number density (bottom right) follows a power-law profile with an exponent of $~-1.5$ (top right).}
\label{fig:plot1}
\end{figure*}

The SDSS\footnote{\tt http://www.sdss.org/} Data Release 6 (DR6) provides imaging data over 9500~deg$^2$ in five photometric bands. Galaxy spectra are determined by CCD imaging and the SDSS 2.5~m telescope on Apache Point, New Mexico \citep{SDSSDR6}. Over half a million galaxies brighter than Petrosian $r$-magnitude $17.77$ over 7400~deg$^2$ are included in the SDSS data with a redshift accuracy better than 30~km/s.

Due to the complications of calculating modified gravity for non-spherical objects, \cite{Prada2007} restricted their analysis only to red galaxies, the vast majority of which are either elliptical galaxies or are dominated by bulges. We followed a similar strategy, restricting our selection of candidate host galaxies to isolated red galaxies. We also restricted our selection to galaxies with a recession velocity between 3000~km/s and 25000~km/s, which yielded approximately 234,000 galaxies in total.

We began our analysis by obtaining a data set from the SDSS. We obtained sky positions, spectra, and extinction-corrected magnitudes for 687,423 galaxies. We adjusted the data set by accounting for the motion of the solar system. We then processed the result using a C-language program that selected, as candidate hosts, isolated red galaxies with no other galaxy within a projected distance of 1~Mpc and a luminosity more than 25\% that of the candidate host. We then identified as candidate satellites galaxies that were within 1~Mpc of projected distance from the candidate host, and had a line-of-sight redshift velocity of less than 1,500~km/s relative to the candidate host. These candidate satellites were binned by distance.
The computation yielded 3,589 hosts with 8,156 satellites. Of these, 121 satellites (or about 1.5\% of the total) were assigned to multiple hosts; no attempt was made to eliminate these duplicates.

The number density of the candidate satellites (Figure~\ref{fig:plot1}, bottom left) suggests that many of these galaxies are not, in fact, satellites. Indeed, if dim galaxies were distributed completely randomly, with no relation to the candidate host, we would expect a number density that increases linearly with projected radius. The actual number density plot appears to be a distribution with a peak at $\sim$100~kpc, superimposed on just such a linear density profile. Subtracting the linear density profile yields the plot in Figure~\ref{fig:plot1} (bottom right), which is a power law profile with exponent $-1.5$, as shown in Figure~\ref{fig:plot1} (top right). (This corresponds to a parameter of $\gamma\simeq -3.5$ in the Jeans equation, discussed below.)

\section{Satellite galaxy velocity dispersion}
\label{sec:results}

Predictions for modified gravity can be made by solving the Jeans equation, which gives the radial velocity dispersion $\sigma_r^2(r)$ as a function of radial distance. \citet{Prada2007} find that neither Newtonian gravity without nonbaryonic dark matter, nor Modified Newtonian Dynamics (MOND) are compatible with observations. \citet{Angus2008}, however, demonstrated that a suitably chosen anisotropic model and appropriately chosen galaxy masses can be used to achieve a good fit for MOND.

Radial velocity dispersions in a spherically symmetric gravitational field can be computed using the Jeans equation \citep{BT1987}:
\begin{equation}
\label{Jeans}
\frac{d(\nu\sigma_r^2)}{dr}+\frac{2\nu}{r}\beta\sigma_r^2=-\nu\frac{d\Phi}{dr},
\end{equation}
where $\nu$ is the spatial number density of particles, $v_r$ is the radial velocity, $\beta(r)=1-[\sigma_\theta^2(r)+\sigma_\phi^2(r)]/2\sigma_r^2(r)$ is the velocity anisotropy, $\Phi(r)$ is the gravitational potential, and we are using spherical coordinates $r$, $\theta$, $\phi$. We can write Eq.(\ref{Jeans}) in the form
\begin{equation}
\label{Jeans2} \frac{d\sigma_r^2}{dr}+\frac{A\sigma_r^2}{r}=-g(r),
\end{equation}
where $g(r)$ is the gravitational acceleration. Here, we have
\begin{equation}
A=2\beta(r)+\gamma(r),
\end{equation}
where $\gamma(r)=d\ln\nu(r)/d\ln{r}$.

If we assume that the velocity distribution of satellite galaxies is isotropic, $\beta=0$. In general, $\beta$ needs to be neither zero nor constant. The number density of candidate satellites favors a value of $\gamma\simeq -3.5$, corresponding to the observed power-law radial density with exponent $-1.5$.

The observed velocity dispersion is along the observer's line of sight, seen as a function of the projected distance from the host galaxy. Therefore, it is necessary to integrate velocities along the line-of-sight:
\begin{equation}
\sigma_\mathrm{LOS}^2(R)=\frac{\int\limits_0^\infty\left[y^2+(1-\beta)R^2\right]r^{-2}\sigma_r^2(y)\nu(y)~dy}
{\int\limits_0^\infty\nu(y)~dy},
\end{equation}
where $\nu$ is the spatial number density of satellite galaxies as a function of distance from the host galaxy, and $y$ is related to the projected distance $R$ and 3-dimensional distance $r$ by
\begin{equation}
r^2=R^2+y^2.
\end{equation}
Changing integration variables to eliminate $y$, we can express the observed line-of-sight velocity dispersion as a function of projected distance as
\begin{equation}
\sigma_\mathrm{LOS}^2(R)=\frac{\int\limits_R^\infty(r^2-\beta R^2)\sigma_r^2(r)\nu(r)/r\sqrt{r^2-R^2}~dr}
{\int\limits_R^\infty r\nu(r)/\sqrt{r^2-R^2}~dr}.
\label{eq:LOS}
\end{equation}

From the field equations derived from the MOG action, we obtain the modified Newtonian acceleration law for weak gravitational fields~\citep{Moffat2006a,Moffat2007e} of a point source with mass $M$:
\begin{equation}
\label{MOGacc} g_{\rm MOG}(r)=\frac{G_NM}{r^2}\left\{1+\alpha\left[1-e^{-\mu r}\left(1+\mu r\right)\right]\right\},
\end{equation}
where $G_N$ is the Newtonian gravitational constant, while the MOG parameters $\alpha$ and $\mu$ determine the coupling strength of the ``fifth force'' vector $\phi_\mu$ to baryon matter and the range of the force, respectively.

In recent work \citep{Moffat2007e}, we have been able to develop formulae that predict the values of the $\alpha$ and $\mu$ parameters from the source mass, in the form
\begin{eqnarray}
\mu&\simeq&\frac{D}{\sqrt{M}},\label{eq:mu}\\
\alpha&\simeq&\frac{M}{(\sqrt{M}+E)^2}\left(\frac{G_\infty}{G_N}-1\right),\label{eq:alpha}
\end{eqnarray}
where the universal parameters
\begin{eqnarray}
G_\infty&\simeq&20G_N,\\
D&\simeq&6250~M_\odot^{1/2}\mathrm{kpc}^{-1},\\
E&\simeq&25000~M_\odot^{1/2}
\end{eqnarray}
are determined from galaxy rotation curves and cosmological observations \citep{Moffat2007e}.

The MOND acceleration $g_{\rm MOND}$ is given by the solution of the non-linear equation
\begin{equation}
\label{MONDacc} g_{\rm MOND}\mu\left(\frac{|g_{\rm MOND}|}{a_0}\right)=\frac{G_NM(r)}{r^2},
\end{equation}
where $M$ is the mass of only baryons and $a_0=1.2\times 10^{-10}$m/s$^2$. The form of the function $\mu(x)$ originally proposed by Milgrom~\citep{Milgrom1984} is given by $\mu(x)=x/\sqrt{1+x^2}$; however, better fits and better asymptotic behavior are achieved using $\mu(x)=x/(1+x)$~\citep{Prada2007}.

Following in the footsteps of \cite{Prada2007}, we grouped satellite galaxy velocities for host galaxies in two luminosity ranges: $-20.5>M^{(1)}_g>-21.1$, and $-21.1>M^{(2)}_g>-21.6$. The corresponding masses for the host galaxies, calculated by \cite{Prada2007} on the basis of the work of \cite{Bell2001}, are
\begin{eqnarray}
M^{(1)}_*=7.2\times 10^{10}~M_\odot&&(-20.5>M^{(1)}_g>-21.1),\label{Mlo}\nonumber\\
M^{(2)}_*=1.5\times 10^{11}~M_\odot&&(-21.1>M^{(2)}_g>-21.6).\label{Mhi}\label{eq:pops}
\end{eqnarray}

We used these values to obtain two sets of predictions for each theory, using $\beta=0$, $\gamma=-2.5$.

\section{The interloper background}

\begin{figure}
\centering\includegraphics[width=\linewidth]{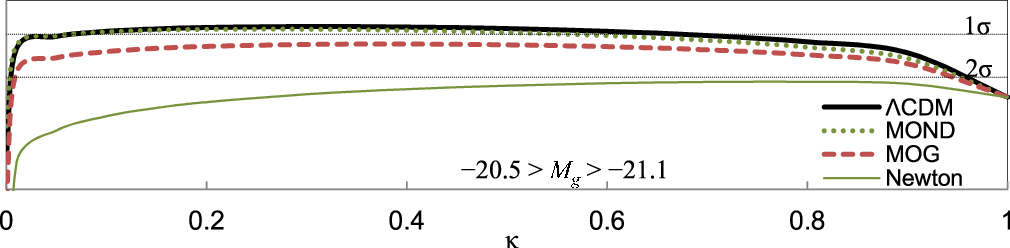}
\centering\includegraphics[width=\linewidth]{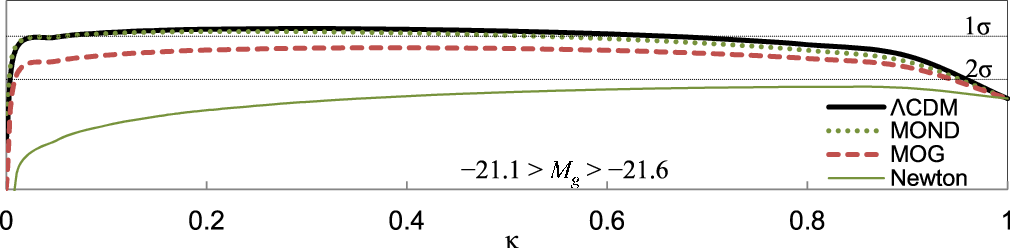}
\caption{Likelihood of $\Lambda$CDM, MOND, MOG, and the Newtonian model without exotic dark matter, as a function of the $\kappa$ parameter as defined in Eq.~(\ref{eq:kappa}). Horizontal lines indicate the 1$\sigma$ and 2$\sigma$ levels relative to the maximum likelihood of the best performing model ($\Lambda$CDM).\vskip 6pt}
\label{fig:plot2}
\end{figure}

Having obtained the velocity dispersion for satellite galaxies around a host galaxy, we now turn our attention to the interloper population.

The actual data consists of host galaxies, satellites, and an effectively random interloper background. When satellites and interlopers are binned by projected distance from host galaxies, the result can be modeled symbolically as
\begin{equation}
N(R)\pm\delta N(R)=N_\mathrm{sat}(R)+N_\mathrm{int}(R),
\end{equation}
where $N(R)$ is the number of galaxies in the bin at projected radius $R$, $N_\mathrm{sat}(R)$ and $N_\mathrm{int}(R)$ are the number of satellites and interlopers, respectively, in that same bin, and $\delta$ is used to represent the sampling error.

This is not the approach taken by \cite{Prada2007}, however. Instead, they elected to subtract a modeled interloper background from the observed number density of satellites, and then compare that to a model representing only satellite galaxies. In effect, they used
\begin{equation}
[N(R)-N_\mathrm{int}(R)]\pm\delta N_\mathrm{sat}(R)=N_\mathrm{sat}(R).
\end{equation}
Assuming that the sampling error of satellites and interlopers are independent, we have
\begin{equation}
\delta N(R)=\sqrt{\delta N^2_\mathrm{sat}(R)+\delta N^2_\mathrm{int}(R)}>\delta N_\mathrm{sat}(R),
\end{equation}
leading to potentially misleading conclusions about the extent to which the galaxy sample can be used to constrain alternate gravity models. It was this realization that led us to repeat some of the analysis performed by \cite{Prada2007}.

For this reason, in our analysis we endeavor to model the actual observation, by estimating both satellite galaxy velocity dispersions in accordance with the previous section and the velocity dispersion of the interloper background. We assume a constant (i.e., independent of distance or sky position) interloper background.

In terms of the polar coordinate $R$ in the sky plane and the line-of-sight velocity $v$, we find that likelihood of finding a satellite between $R$ and $R+dR$, with line-of-sight velocity between $v$ and $v+dv$, will be proportional to
\begin{equation}
p_S\propto R^{\gamma+2}\exp\left(\frac{-v^2}{2\sigma^2_\mathrm{LOS}(R)}\right),
\end{equation}
subject to normalization to ensure that the probability of finding a particular satellite somewhere within the observational range ($0\le R\le 1$~Mpc, $0\le v\le 1500$~km/s) is unity. On the other hand, the likelihood of finding an interloper from a uniformly distributed background, between $R$ and $R+dR$, is
\begin{equation}
p_I\propto R,
\end{equation}
again subject to normalization. If we assume that the proportion of interlopers is $\kappa$ ($0\le\kappa\le 1$), the combined probability of finding a galaxy (satellite or interloper) at $R, v$, is
\begin{equation}
p=\kappa p_I + (1-\kappa)p_S,\label{eq:kappa}
\end{equation}
We can use this value of $p$ to develop the likelihood function $L(\kappa)=\prod{p(\kappa)}$ for the two candidate satellite populations (\ref{eq:pops}), choosing the value of $\kappa$ to obtain the maximum likelihood.

Using this likelihood function, we find that $\Lambda$CDM is the best performing model, marginally outperforming MOND and MOG, with maximum likelihood obtained at $\kappa=0.313^{+0.300}_{-0.194}$ and $\kappa=0.295^{+0.295}_{-0.183}$, respectively, for the two candidate satellite populations. The $\Lambda$CDM and MOND models are effectively indistinguishable (see Figure~\ref{fig:plot2}). They both outperform MOG, but the difference is not statistically significant: comparison with a $t$-statistic yields a probability of 24.2\% (for $-20.5>M_g>-21.1$) and 23.0\% (for $-21.1>M_g>-21.6$) that the difference between MOG and $\Lambda$CDM is due to chance. Only the Newtonian prediction without nonbaryonic dark matter can be excluded with a $2\sigma$ significance.

For this reason, it seems futile to use this type of statistical analysis of satellite galaxies to distinguish between CDM models on the one hand, and various gravitational theories on the other, due to the presence of the interloper population.

\section{Conclusions}
\label{sec:conclusions}

Observational data presented by \cite{SDSSDR6} and studied by \cite{Prada2007} are viewed as evidence of the success of the $\Lambda$CDM model. The usual approach relies on the critical step of interloper removal, before the data is compared against predictions. We argue that this approach is fundamentally flawed: rather than attempting to remove interlopers from the data, we must add the interloper background theoretical predictions, in order to predict observational values. When we carry out this approach, we find the $\Lambda$CDM and modified gravity theory predictions cannot be distinguished and that although the SDSS data set weakly favors $\Lambda$CDM over the alternatives, it cannot be used to falsify any of the theories we examined.

~

\section*{Acknowledgments}

The research was partially supported by National Research Council of Canada. Research at the Perimeter Institute for Theoretical Physics is supported by the Government of Canada through NSERC and by the Province of Ontario through the Ministry of Research and Innovation (MRI).

~\par ~\par

\bibliography{refs}
\bibliographystyle{apj}

\end{document}